\documentclass[twocolumn,prb,preprintnumbers,amsmath,amssymb,citeautoscript,nobalancelastpage,showpacs]{revtex4}

\usepackage{graphicx}
\usepackage{dcolumn}
\usepackage{bm}
\usepackage{times}
\usepackage{color}
\usepackage{setspace}

\begin{document}
\title{X-ray crystal structure analysis and the Ru valence of
Ba$_{4}$Ru$_{3}$O$_{10}$ single crystals}
\author{Taichi Igarashi$^{1}$}
\email[Electronic address: ]{igarashi.taichi@b.mbox.nagoya-u.ac.jp}
\author{Yoshio Nogami$^{2}$}
\altaffiliation{Visiting professor in the Department of
Physics, Nagoya University}
\author{Yannick Klein$^{3}$}
\author{Gwenaelle Rousse$^{3}$} 
\author{Ryuji Okazaki$^{1}$}
\author{Hiroki Taniguchi$^{1}$}
\author{Yukio Yasui$^{4}$} 
\author{Ichiro Terasaki$^{1}$}
\affiliation{$^{1}$Department of Physics, Nagoya University, Nagoya 464-8602, Japan}
\affiliation{$^{2}$Department of Physics, Okayama University, Okayama 700-8530, Japan}
\affiliation{$^{3}$Universit\'e Pierre et Marie Curie-Paris 6, IMPMC-CNRS
UMR 7590 Campus
Jussieu, 4 place Jussieu F-75252 Paris Cedex 05, France}
\affiliation{$^{4}$Department of Physics, Meiji University, Kawasaki
214-8571, Japan}
\begin{abstract}
We present the single-crystalline x-ray diffraction study on the
   Ba$_{4}$Ru$_{3}$O$_{10}$ consisting of the
corner-shared Ru$_{3}$O$_{12}$ trimers. The crystal structure is
  re-determined from 78 to 300 K across an
  antiferromagnetic transition at 105 K.
 The orthorhombic symmetry ($Cmca$, space group No. 64) is preserved at
  all temperatures measured.
This structure presents exceptionally long Ru-O distances characterized
  by a significant distribution within the Ru$_{3}$O$_{12}$ trimer.
 A bond valence sum calculation suggests that the charge disproportionation
 within the Ru$_{3}$O$_{12}$ trimer emerges even at room temperature,
 which we ascribe to molecular
 orbital formation in the Ru$_{3}$O$_{12}$ trimer, as supported by
 recent theoretical calculations.
Based on the analyzed crystal structure, the electronic states and the
nature of the phase transition at 105 K are discussed.
\end{abstract}

\maketitle
\section{Introduction}
\ \ Ruthenium oxides including tetravalent Ru ions exhibit various
fascinating electronic and magnetic states owing to the multiple degrees
of freedom of 4$d$ electrons as seen in complex electronic phase
diagrams of $A$RuO$_{3}$ and $A_{2}$RuO$_{4}$ ($A$ = Ca, Sr).
In Ca$_{2-x}$Sr$_{x}$RuO$_{4}$, for instance, the ground state varies
from an antiferromagnetic insulating\cite{Nakatsuji} ($x < 0.2$) to a superconducting
phase\cite{YMaeno2} ($x = 2$) through a spin-glass state in the wide
composition range\cite{JPCarlo}. 
Structurally, these compounds belong to the Ruddlesden-Popper phases where the 
RuO$_{6}$ octahedra are corner-shared. In contrast, when Ba$^{2+}$ is
combined with Ru$^{4+}$,
the ruthenates derive from the hexagonal perovskite-type structure. 
An important feature in
the hexagonal ruthenates is that the
RuO$_{6}$ octahedra are face-shared, leading to a shorter Ru-Ru distance
than in Ru metal. This indicates a stronger hybridization of the Ru
4$d$ orbitals and the resulting molecular orbital formation may introduce
an additional internal charge degree of freedom in such Ru multimer,
which can be an origin for exotic electronic properties in this system.
\par
 \begin{figure}[htbp]
 \begin{center}
\includegraphics[width=80mm]{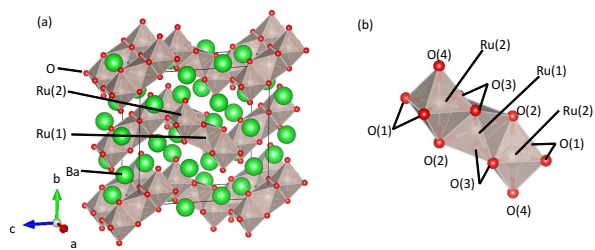}
 \end{center}
\label{structure}
\caption{(Color online) (a) The crystal structure of Ba$_{4}$Ru$_{3}$O$_{10}$. (b)
  Ru$_{3}$O$_{12}$ trimer in Ba$_{4}$Ru$_{3}$O$_{10}$.}
 \end{figure}
\ \ The barium ruthenate
Ba$_{4}$Ru$_{3}$O$_{10}$ with tetravalent Ru ions\cite{CDussarrat} is related to 
the hexagonal-perovskite 9R-BaRuO$_{3}$ in which Ru$_{3}$O$_{12}$ trimers made of face shared
RuO$_{6}$ octahedra. While in the latter compound, 
each trimer is connected with six neighboring trimers 
 via the corners of the outer RuO$_{6}$ octahedra, the number of
 connections is limited to four in Ba$_{4}$Ru$_{3}$O$_{10}$, and This compound
corrugated layers stacked along the $b$ axis [Fig. 1(a)]. Each trimer is built upon
 two inequivalent Ru sites, where the center and outer Ru
 ions are denoted as Ru(1) and Ru(2), respectively [Fig. 1(b)]. The structure was
 reported to be orthorhombic with space group $Cmca$ (No.64) 
at room temperature by Carim $et$ $al.$\cite{AHCarim}, and no
 symmetry lowering was detected down to 10 K in a neutron powder diffraction
 experiment by Klein $et$ $al.$\cite{YKlein} An intriguing feature in
 Ba$_{4}$Ru$_{3}$O$_{10}$  is an unusual phase
 transition at $T_{N} = 105$ K, below which an antiferromagnetic order
 develops with an opening of an energy gap in the charge transport phenomena\cite{YKlein}.
This antiferromagnetic order is unique in the sense that it does not
break the translational symmetry of the lattice, but is characterized by a sudden drop
from the susceptibility experiment. Moreover, neutron powder diffraction
patterns indicate the growth of magnetic intensity on the nuclear (0
0 2) reflection below $T_{N}$.
\par
\ \ Here we present the synthesis and x-ray structural analysis of
Ba$_{4}$Ru$_{3}$O$_{10}$ single crystals in a wide temperature range
from 78 to 300 K with a reliability factor ($R$-factor) better than 2\%. All the
low-temperature studies of the title
compound reported thus far were done with polycrystalline samples.
Owing to the complicated structure, the Rietveld refinements do not give
satisfactory reliability factors. In contrast, single crystal structure analysis
can make full use of all the Bragg reflections on the Ewald sphere, and
the structure can be solved with a model-free approach.
We have confirmed the space group to be $Cmca$ using
extinction rule at all temperatures above 78 K, and have found no 
additional superlattice reflections below
$T_{N}$, which indicates no structural symmetry
breaking in this transition. 
On the basis of the determined structure, we have further found that
some bond lengths are anomalously long, which exhibit substantial
temperature dependence.
Using a bond valence sum
calculation, we evaluate the formal valences of Ru(1) and Ru(2) as a
function of temperature, and discuss the electronic states and the
nature of the 105 K transition.
\section{Experiments}
\ \ Single crystals of Ba$_{4}$Ru$_{3}$O$_{10}$ were synthesized
from stoichiometric mixture of BaCO$_{3}$ (99.9\%) and RuO$_{2}$ (99.9\%) using a
solid state reaction. The stating
materials were mixed in an agate mortar with a pestle, and heated in an
alumina crucible at $1000\ {}^\circ\mathrm{C}$ for 12 h in air. The
obtained powder was mixed
and heated again at $1200\ {}^\circ\mathrm{C}$ for 24 h in air. 
The powder was then reground, pressed into
pellets, and heated at $1400\ {}^\circ\mathrm{C}$ for 24 h in air. The pellets
were annealed at $1400\ {}^\circ\mathrm{C}$ for 192 h, and small single
crystalline samples were grown on the surface of the pellets.
\par
\ \ In order to determine the crystal structure precisely we
measured two samples (Sample1 and Sample2) in different conditions. The
dimensions of Sample1 and Sample2 were approximately 
$40\times 50\times 150$ $\mathrm{\mu}$m$^{3}$ and 
$50\times 60\times 160$ $\mathrm{\mu}$m$^{3}$, respectively.
Sample1 was used to determine the space group and the unit cell
parameters from 300 down to 78 K.
The structural determination was carried out with
an automated RIGAKU Saturn Varimax system equipped with 
 CCD
detector. The instrument employed Mo$K\alpha$ radiation at 50 kV and 24 mA using 
a doubly focused mirror. The sample stage was controlled by a triaxial rotation system
($\omega, \chi, \phi$) equipped with 1/4 $\mathrm{\chi}$.
Sample1 was measured with the oscillation method, where 
the rotation parameters were fixed to 
$2\theta = 20$ deg, $\chi = 45$ deg, and $\phi = 0$ deg, and
 $\omega$ was swept from 0 to 180 deg with an oscillation angle
 $\delta = 0.5$ deg during a counting time of 1 s.
Sample2 was also measured by the oscillation method, but the
 measurement was expanded to 4-times larger range on the Ewald
sphere to determine the atomic coordinates more precisely.
 Specifically, $\omega$ was swept from 0 to 180 deg under the conditions of
$(2\theta,\phi)$ = (20 deg, 0 deg), (20 deg, 90 deg),
(70 deg, 180 deg), and (70 deg, 270 deg). The temperature was controlled with spraying
nitrogen gas, and was monitored at the outlet of the gas. We use this
temperature throughout the present manuscript, which is approximately
3 K lower than the sample temperature. The space group and the lattice
constants were computationally determined with Crystal Clear,
RIGAKU, and the integrated intensity for all the reflections was
calculated simultaneously. The crystal structure was determined from the
obtained reflection intensity using a direct method
with SHELX-97\cite{Shelx} and Yadokari-XG 2009\cite{Yadokari}.
\section{Results and Discussion}
\ \ From a careful investigation of the Laue symmetry and the extinction
rule, the structural symmetry at 300 K is determined to be
orthorhombic ($Cmca$), with lattice parameters of $a$ = 5.7740(5) \AA,
$b$ = 13.2571(10) \AA, and c = 13.0649(9) \AA\ and the structure determined on
single-crystalline samples by Carim $et$ $al$. is confirmed. 
In order to investigate the symmetry lowering below $T_{N}$, 
we have examined the extinction rule for the Bragg reflections at 78 K,
and have found no superlattice reflections stronger than $10^{-5}$ of the fundamental 
reflection intensity.
This indicates that the $Cmca$ symmetry holds below $T_{N}$, and that
the phase transition is not driven by lattice instability. We have
carefully examined the temperature dependence of the (0 0 2) intensity,
and have found it essentially independent of temperature below
$T_{N}$. Thus we can conclude that the growth of the (0 0 2) intensity
observed in the neutron experiment below $T_{N}$ is of magnetic origin,
as Klein $et$ $al$.
suggested previously\cite{YKlein}.
The atomic coordinates and isotropic thermal parameters are listed in
Table I, which agree with the previous works\cite{YKlein, AHCarim}.
\begin{table}[htbp]
\caption{Atomic coordinates and thermal parameters for
 Ba$_{4}$Ru$_{3}$O$_{10}$ at 300 K.}
\label{t1}
\begin{tabular}{lllllll}
\hline
 Atom & Site &\ \ \ \ & $x$ & $y$ & $z$ & $\beta$ (\AA$^{2}$)\\
\hline 
Ba(1) & 8f && 0 & 0.23970(3) & 0.11143(3) & 0.0106(3) \\
Ba(2) & 8f && 0 & 0.53555(3) & 0.13884(3) & 0.0082(3) \\
Ru(1) & 4a && 0 & 0 & 0 & 0.0058(3) \\
Ru(2) & 8f && 0 & 0.87538(4) & 0.14943(4) &0.0052(3) \\
O(1) & 8e && 1/4 & 0.3784(4) & 1/4 & 0.0118(13) \\
O(2) & 8f && 0 & 0.0360(4) & 0.1521(4) & 0.0061(12) \\
O(3) & 16g && 0.2736(6) & 0.3901(2) & 0.0347(3) & 0.0072(8) \\
O(4) & 8f && 0 & 0.7285(5) & 0.1482(4)& 0.0141(14) \\
\hline
\end{tabular}
\end{table}
\par
\ \ Table I\hspace{-.1em}I lists the temperature dependence of
the lattice parameters, reliability
factors, and the Ru-O and Ru(1)-Ru(2) distances for Sample1 and Sample2. 
All the refinements lead to reliability factors $R_{1}$ (all data) 
with a precision better than 0.02.
At 300 K, the Ru(1)-Ru(2) distance is determined to be 2.5576(6)
\AA. This is significantly shorter than that of the ruthenium metal (2.65 \AA), and
decreases with decreasing temperature, which
implies that the hybridization of the Ru 4$d$ orbitals becomes stronger at
low temperatures.
\begin{table*}[htbp]
\caption{Temperature dependence of structure analysis data and selected
  inter atomic distances for Sample1 and Sample2.}
\label{t2}
\center
\includegraphics[width=0.8\linewidth]{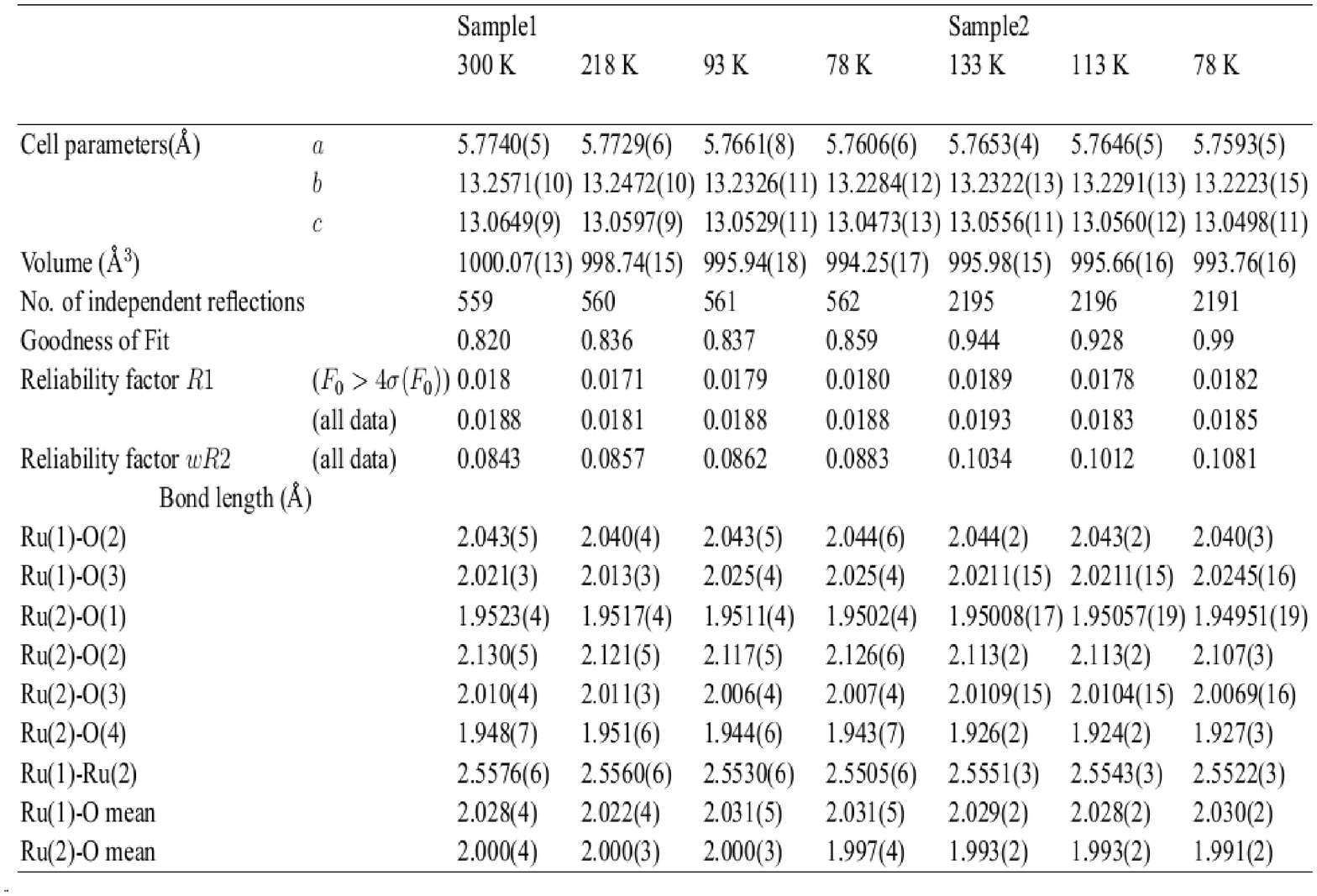}
\end{table*}
\par
\begin{figure}[htbp]
 \begin{center}
\includegraphics[width=80mm]{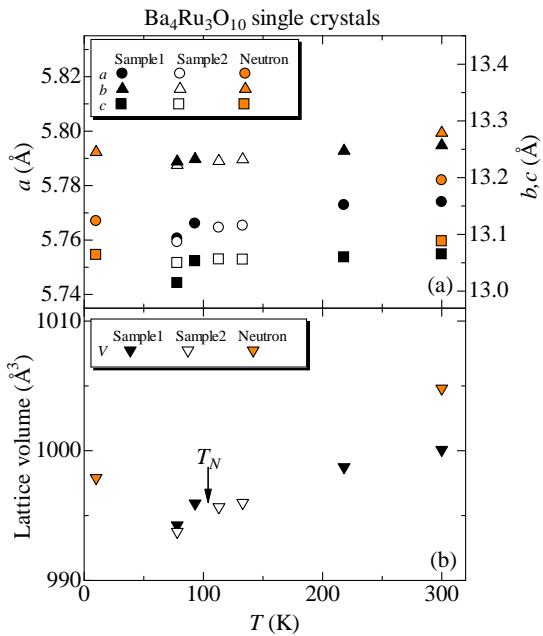}
 \end{center}
\label{volume}
\caption{(Color online) (a) The lattice constant $a$ (left scale), $b$, and $c$ (right
 scale) as a function of temperature. (b) The unit cell volume of
 Ba$_{4}$Ru$_{3}$O$_{10}$ as a function of temperature. Neutron data on
 the polycrystalline samples are also shown\cite{YKlein}.}
 \end{figure}
\ \ Figure 2 shows the temperature dependence
of the cell parameters using the data for Sample1 and Sample2 together
with the neutron data in Ref. 6, where all of the cell parameters
decrease with decreasing temperature. The $b$ axis decreases by 0.2 \%
between 300 and 78 K for Sample1, which is larger than the change in the
other axes. The parameters for Sample1 and Sample2
are 1.7 \% smaller
than the neutron data, which could come from delicate difference in
chemical composition between the polycrystalline sample and single
crystals and in the measurements between x-ray structural analysis and
neutron powder diffraction. The lattice volume differs only by 0.05 \% between Sample1 and
Sample2 at 78 K.
\par
\begin{figure}[htbp]
 \begin{center}
\includegraphics[width=70mm]{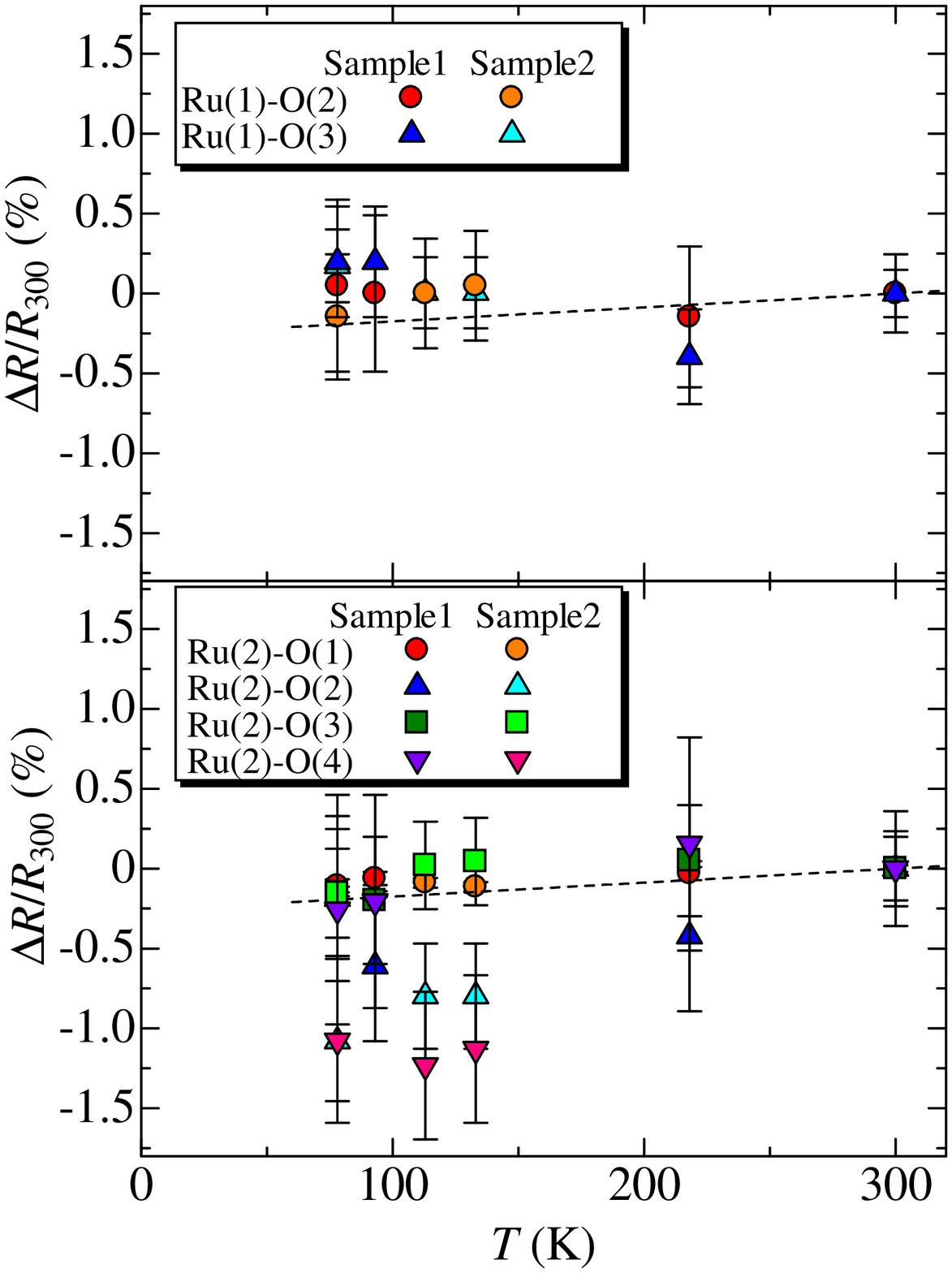}
 \end{center}
\label{dist}
\caption{(Color online) The relative change in the Ru-O distances of
 Ba$_{4}$Ru$_{3}$O$_{10}$ as a function of temperature. Dashed lines are
 the volume shrinkage from 300 K. $\Delta R/R_{300}$ is referred to
 the text.}
 \end{figure}
\ \ Figure 3 shows the temperature dependence of the relative change in the
Ru-O distances $\Delta R/R_{300}$ together with  the temperature evolution
of volume shrinkage plotted using the dashed lines.
Here, $\Delta R$ is a change of the Ru-O distance from 300 K
and $R_{300}$ is the Ru-O distance at 300 K.
The Ru(1)-O(2) and Ru(1)-O(3) values roughly lie on the dashed line.
In contrast, the Ru(2)-O distances plunge under the dashed line,
and in particular, the Ru(2)-O(2) and Ru(2)-O(4) distances for Sample2
significantly decrease below 150 K.
This indicates that the volume of the Ru(2)O$_{6}$ octahedron shrinks more
remarkably than the thermal average, while that of the Ru(1)O$_{6}$
octahedron is weakly dependent on temperature.
\par
We should emphasize that the mean Ru-O distance of Ba$_{4}$Ru$_{3}$O$_{10}$
is exceptionally long. As listed in Table I\hspace{-.1em}I, the
mean distances of Ru(1)-O and Ru(2)-O are 2.028(4)
and 2.000(4) \AA \ at 300 K, respectively. These values are significantly larger than the mean
Ru-O distance in Sr$_{2}$RuO$_{4}$ and Ca$_{2}$RuO$_{4}$ (1.98-1.99
\AA)\cite{MBraden}, which implies that the formal Ru valence in the
title compound is lower than 4+ on the basis of a bond-valence sum
calculation\cite{BVS_para}.
We should further note that
Ru(1)-O is longer than Ru(2)-O, meaning that the formal valence of Ru(1)
is lower than that of Ru(2).
\par
\begin{figure}[htbp]
 \begin{center}
\includegraphics[width=90mm]{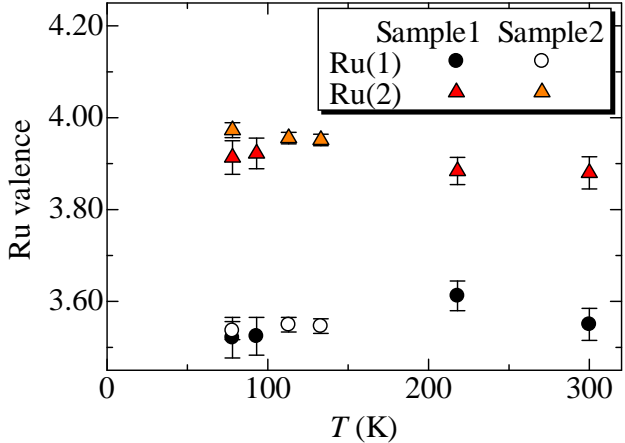}
 \end{center}
\label{BVS}
\caption{(Color online) The Ru(1) and the Ru(2) valences of
 Ba$_{4}$Ru$_{3}$O$_{10}$ as a function of temperature.}
 \end{figure}
\ \ Based on the analyzed structure, we
evaluate the formal Ru valence using the bond valence sum
method, where we use 1.834 \AA \ as a standard distance $d_{0}$ of Ru$^{4+}$
ion\cite{BVS_para}.
 The formal valences of Ru(1) and Ru(2) are calculated to be 3.55 and 3.88 at 300 K,
respectively, as shown in Fig. 4, as was already expected from the long Ru-O distances discussed
above. 
A valence lower than 4+ is directly related to the longer Ru-O
distances, possibly because the
intra-trimer Ru ions push the intervening oxygen ions away by coming
closer to each other.
 More importantly, the valence difference of 0.33 between the inner and
 outer Ru
is far larger than what is commonly observed in the charge-ordered oxides, e.g.,
Pr$_{0.6}$Ca$_{0.4}$MnO$_{3}$\cite{DAladine} and
Fe$_{3}$O$_{4}$\cite{MSSenn}. This
implies that the degree of the charge disproportionation can be much larger,
meaning that it may exceed +$e$. If different $d_{0}$ distances are
assumed for the Ru(1) and Ru(2) sites as was done for
LiMn$_{2}$O$_{4}$\cite{Rodriguez}, the valences of Ru(1) and Ru(2) can
be evaluated to be 2.99+ ($d_{0} = 1.77$ \AA\ for Ru$^{3+}$) and 4.64+
($d_{0} = 1.90$ \AA\ for Ru$^{5+}$)
respectively\cite{Ru3+}.
These valences agree with our proposed model based on a localized
picture of $t_{2g}$ electrons (See the discussion below).
\par
\ \ Let us consider the electronic
states of the Ru$_{3}$O$_{12}$ trimer. Very recently, two groups were
independently calculated the electronic sates of
Ba$_{4}$Ru$_{3}$O$_{10}$\cite{Streltsov, Radtke}The two papers discussed the
electronic states of the title compound from an itinerant picture using
the density functional theory. 
We will start from the electronic
states of an isolated Ru$_{3}$O$_{12}$ trimer, and then introduce a
small inter-trimer transfer. 
This approach may not be exact, but can give us intuitive insight for
the electronic states and the electronic transition of
Ba$_{4}$Ru$_{3}$O$_{10}$.
As shown in the left side of Fig. 5(a), the four $d$ electrons in the
Ru$^{4+}$ ion partially occupy the
$t_{2g}$ orbitals to make the low-spin state. 
At first, we consider that the octahedra in each Ru site are
compressed along the Ru-Ru direction, and $t_{2g}$ orbitals split into
$a_{1g}$ and $e_{g}'$ orbitals with a crystal-field splitting of
$\Delta$ as is schematically shown in the right
side of Fig. 5 (a). 
We then consider the formation of molecular orbital for all the orbitals. 
We set the quantization z axis to be parallel to the Ru-Ru
direction. Since one of the orbitals in each Ru site is extended to the
z axis at a 180-degree angle of Ru(2)-Ru(1)-Ru(2), the $a_{1g}$ orbitals
form a strong $\sigma$ bonding. 
By neglecting the correlation effects, we can
construct bonding (B), non-bonding (NB) and anti-bonding (AB) orbitals.
We also take account of the $e_{g}'$ orbitals with $\delta$ bonding.
\par
\ \ Figure 5(b) shows the electronic states in three Ru ions without
Ru-Ru transfer energies. 
 \begin{figure}[htbp]
 \begin{center}
\includegraphics[width=80mm]{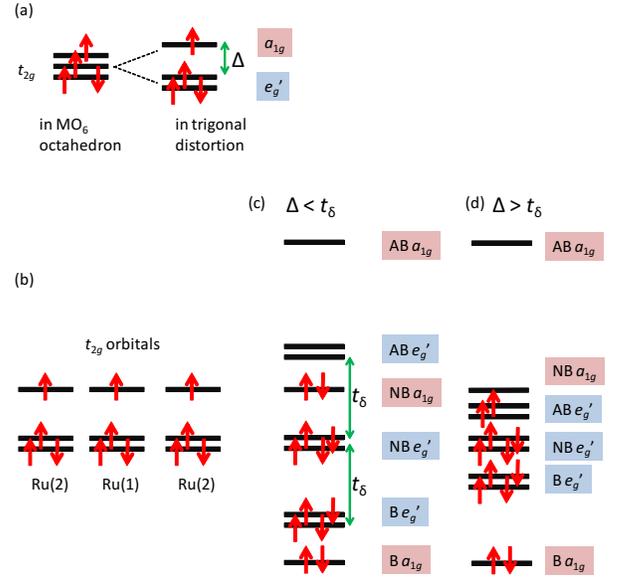}
\caption{(Color online) 
(a) Splitting $t_{2g}$ orbital in $a_{1g}$ and $e_{g}'$ orbitals
  caused by trigonal distortion in a  trimer. 
  (b) $t_{2g}$ orbitals in the three Ru sites. The energy levels in the
  trimer (c) for $\Delta < t_{\delta}$ and (d) for 
$\Delta > t_{\delta}$. The $a_{1g}$ and $e_{g}'$ orbitals
  hybridize to form the anti-bonding (AB),
  non-bonding (NB) and bonding (B) orbitals.}
 \end{center}
 \end{figure}
We use the simplest
Hamiltonian given by\cite{Streltsov, Radtke}
\begin{equation}
H = \begin{bmatrix}
     0 & t_{i} & 0 \\
     t_{i} & 0 & t_{i} \\
     0 & t_{i} & 0 \\
    \end{bmatrix}
(i = \sigma, \delta).
\end{equation}
Then molecular orbital is expressed by 
$\Psi_{AB} = (\phi_{1}-\sqrt{2}\phi_{2}+\phi_{3})/2$, 
$\Psi_{NB} = (\phi_{1}-\phi_{3})/\sqrt{2}$, and
$\Psi_B = (\phi_{1}+\sqrt{2}\phi_{2}+\phi_{3})/2$, where $\phi_{1}$, $\phi_{2}$, and
$\phi_{3}$ are the orbitals in the left Ru(2), Ru(1), and right Ru(2) ions,
respectively. 
The energy splitting is given as $\pm t_{\sigma}$ for the $a_{1g}$
orbital, and $\pm t_{\delta}$ for the $e_{g}'$ orbital.
\par
\ \ 
Let us first consider $\Delta < t_{\delta}$.
The highest
occupied orbital is the fully-filled NB $a_{1g}$ orbital. This
electronic configuration is, however, nonmagnetic, which is in a serious
contradiction to the experimental fact that Ru(2) is
magnetic\cite{YKlein}. The valence of Ru(1) and Ru(2) site are 5+ and 3.5+,
because the NB orbitals have no weight at the Ru(1) site, as was
indicated in the previous works \cite{Streltsov, Radtke, Bursten}. 
This estimation also disagrees with the BVS calculation result that the Ru(1)
is of lower valence. 
When $\Delta > t_{\delta}$, the highest occupied state is AB $e_{g}'$,
where the spins of the two electrons are aligned in parallel. In this
case, however, the magnetic moment exists in the Ru(1) site, which disagrees
with the neutron experiment.
\color{black}
\par \ \ 
 \begin{figure}[htbp]
 \begin{center}
\includegraphics[width=80mm]{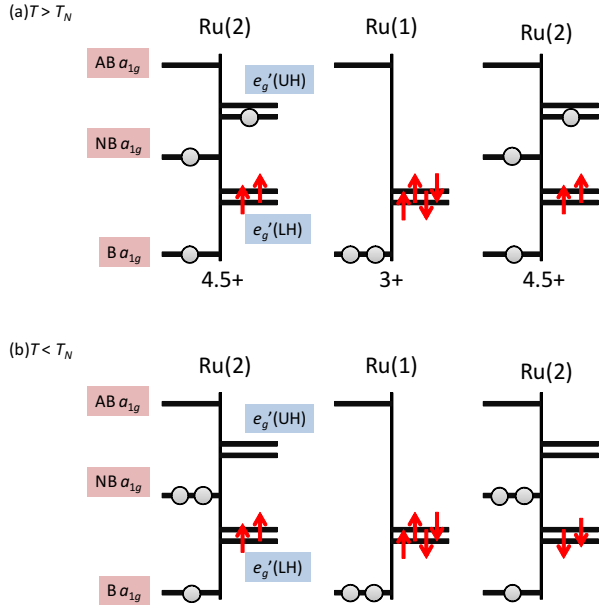}
\label{fig.5}
\caption{
(Color online) Proposed energy diagram for the Ru trimer in
  Ba$_{4}$Ru$_{3}$O$_{10}$. (a) $T > T_{N}$ and (b) $T < T_{N}$. The
  energy levels for the molecular orbitals of the $a_{1g}$ orbitals are
  drawn in the left, and those for the $e_{g}'$ orbitals in the right. A
  grey circle represents 0.5 electron and an arrow represents a
  localized electron. (See text)
\color{black}}
 \end{center}
 \end{figure}
In order to modify the model
in Fig. 5, we retain the B, NB and AB $a_{1g}$ orbitals, but
introduce the upper Hubbard (UH) and lower Hubbard (LH) $e_{g}'$
orbitals. 
This is based on the fact that the magnetic susceptibility of Ba$_{4}$Ru$_{3}$O$_{10}$ above
$T_{N}$ suggests an existence of a local moment. We also note that
similar local moments are suggested in some Ru oxides such as
Ba$_{3}$CoRu$_{2}$O$_{9}$\cite{HDZhou} and
Ba$_{3}$BiRu$_{2}$O$_{9}$\cite{WMiiller}, where $a_{1g}$ and $e_{g}'$
orbitals can act different roles.\cite{SVStreltsov}.
To express the $a_{1g}$ molecular orbitals and the Hubbard-split
$e_{g}'$ orbitals simultaneously, we draw energy levels in an
unconventional way in Fig. 6. As shown in Fig. 6, each Ru ion shows the
$a_{1g}$ levels in the left and the $e_{g}'$ levels in the right.
While the energy diagram for the $e_{g}'$ levels follows the convention,
that for the $a_{1g}$ levels represents the weight for
the wave functions of $\phi_{1}$, $\phi_{2}$, and $\phi_{3}$.
Let us focus the bonding $a_{1g}$ orbital, which is expressed as 
$(\phi_{1}+\sqrt{2}\phi_{2}+\phi_{3})/2$. Thus the weight for the 1, 2,
3 sites are 0.5e, 1.0e and 0.5e, respectively. We denote this situation
by using gray circles in Fig. 6. Similarly, the half-filled NB $a_{1g}$
orbital is expressed as $(\phi_{1}-\phi_{3})/\sqrt{2}$, which
corresponds to 0.5e in the Ru(2) sites, as shown in Fig. 6(a).
\color{black}
\par
\ \ To understand the charge disproportionation within the trimer, let us
count the number of electrons in the Ru(1) and Ru(2) ions.
As discussed above, the $a_{1g}$ molecular orbitals give 1.0e
for each Ru site.
\color{black}
Owing to the deep level of
the bonding $a_{1g}$ orbital, the center of gravity in the density of
states for Ru(1) shifts to lower energies, which implies that a finite
charge should be transfered from Ru(2) to Ru(1)\cite{Streltsov}.
Assuming that a charge of 0.5e is transfered from Ru(2) UH $e_{g}'$ orbitals 
to Ru(1) $e_{g}'$ orbital (i.e.,
Ru(1) is 3+ and Ru(2) is 4.5+),
the four electrons fully occupy the
$e_{g}'$ levels in the Ru(1) site to let the Ru(1) ion be nonmagnetic as shown in
Fig. 6(a). 
On the other hand, 0.5e, 0.5e and 2.0e occupy the levels of
NB $a_{1g}$, UH $e_{g}'$ and LH $e_{g}'$ in the Ru(2) site,
respectively.
At first sight, the highest occupied level of the UH $e_{g}'$ orbitals
in Fig. 6(a) is unstable against the half-filled NB $a_{1g}$ orbital. We
think that this instability will be removed by introducing an inter-trimer hopping
to form the $a_{1g}$ and $e_{g}'$ bands.
\color{black}
Figure 6(a) captures some essential features given in Refs. 13 and
14. The charge disproportionation in Ref. 13 is comparable with ours,
and the charge distribution between $a_{1g}$ and $e_{g}'$ in Ref. 14 is
similar to ours.
\par
\ \ Based on this self-disproportionation  picture, 
we can explain the following experimental facts from the
above mentioned intra-trimer charge disproportionation of
Ru$^{3+}$/Ru$^{4.5+}$:
(i) The lower Ru(1) valence reasonably close to 3+, 
(i\hspace{-.1em}i) The valence difference larger than that in
Mn$^{3+}$/Mn$^{4+}$ ordered materials suggested from the
bond-valence-sum calculation,
(i\hspace{-.1em}i\hspace{-.1em}i) the nonmagnetic Ru(1) site and the
magnetic Ru(2) site, (i\hspace{-.1em}v) no charge gap above $T_{N}$, (v)
 nonmetallic resistivity above $T_{N}$ implying
narrow conduction bands formed by the NB $a_{1g}$ and the localized UH
$e_{g}'$ bands, and (v\hspace{-.1em}i) 
the small positive thermopower and negative Hall coefficient suggesting 
 multiband conduction.
However, it is fair to point out some open issues.
One is how to directly observe the large difference of valence between Ru(1) and
Ru(2) site. 
Second one is how to observe the Hubbard-split $e_{g}'$
 orbitals. Optical or photoemission measurements are needed to explore
 the electronic states.    
\color{black}
\par
\ \ We have further found that the charge disproportionation develops with
decreasing temperature. Figure 4 shows the formal Ru valence plotted as
a function of temperature. Mirroring the weak temperature dependence of
the Ru(1)-O distances, the Ru(1) valence is nearly independent of
temperature. In contrast, the significant shrinkage of the Ru(2)-O(2)
and Ru(2)-O(4) distances results in the increase of the Ru(2) valence
from 300 down to 78 K. These results indicate that the charge
disproportionation in this compound does not arise simply from a band
picture, but from some cooperative motion of 4$d$ electrons. Although the
valence change is so smooth below and above $T_{N}$, the valence
difference at 78 K reaches 0.45, which is anomalously large in comparison
with other charge-ordered materials. 
\par
\ \ We do not yet understand the mechanism of the 105 K transition from the
paramagnetic metal to the antiferromagnetic insulator, but will try to
discuss some possible scenarios on the basis of the electronic states
discussed above. We can exclude possibilities of charge and spin
density waves, because these phases accompany the translational symmetry
breaking\cite{Gruner}.
For the same reason, we can also exclude a possibility of the Slater
insulator\cite{YGShi}.
One possibility is driven by the short-range magnetic order.
In the title compound, a broad maximum around
200 K in the susceptibility implies that a short-range antiferromagnetic
correlation develops below 200 K, which has been analyzed in terms of spin dimer above $T_{N}$
by Radtke $et\ al.$\cite{Radtke}.
We propose a possible electron configuration below $T_{N}$ in Fig. 6(b).
Assuming the Hund coupling between the NB $a_{1g}$ and the $e_{g}'$
electrons, we expect that the electron spin in the NB $a_{1g}$ orbital
tends to polarize in parallel to
the spin in the $e_{g}'$ orbitals. 
However, since the spins of the Ru(2) sites are ordered in antiparallel,
the spin in the NB $a_{1g}$ orbital is frustrated. 
To relieve this frustration, we propose that a charge of 0.5e will be
transfered from the UH $e_{g}'$
orbitals to the NB $a_{1g}$ orbital to let the NB $a_{1g}$ be
nonmagnetic with opening the charge gap between the NB $a_{1g}$ orbital
and the UH $e_{g}'$ orbitals.
In this respect, this electronic phase transition can be viewed as an
orbital ordering transition.
\par
\ \ Finally, we should emphasize that the Ru valence
difference observed here is intrinsic to the hexagonal-perovskite
ruthenates.
In the related ruthenate BaRuO$_{3}$ composed of similar trimers\cite{PCDonohue}, the
calculated partial density of states\cite{CFelser} and the neutron diffraction
\cite{ASantoro} have revealed that the valence of the Ru(1) site is lower than 4+.
A vitally important difference is that a phase transition does not
occur, but a pseudogap is observed in the optical conductivity and the
charge transport in BaRuO$_{3}$\cite{TWNoh, YKleinBRO3}. 
We think that the origin of the pseudogap is the intra-trimer charge
disproportionation discussed in this article, and that the absence of the phase transition should
be ascribed to the different trimer network from 
Ba$_{4}$Ru$_{3}$O$_{10}$.

\section{Summary}
\ \ We have successfully grown high-quality Ba$_{4}$Ru$_{3}$O$_{10}$ single
crystals, and have solved the crystal structure with a reliability
factor better than 2\% at all the temperatures from 78 to 300 K through
x-ray crystal analysis. This ruthenate is orthorhombic with space group
$Cmca$ (space group No. 64) at all temperatures, and no symmetry
breaking is detected below the antiferromagnetic transition of 105 K.
We have further found that the Ru-O distances are anomalously long, and
that the formal Ru valence is smaller than 4+ according th the
bond-valence-sum calculation. Most remarkably, a large charge
disproportionation of 0.45$e$ within the Ru$_{3}$O$_{12}$ trimer is
obtained at 78 K, which is ascribed to the strong hybridization of the Ru $a_{1g}$
orbital. We have explained the experimental results on the basis of the
solved structure, and proposed a model from a localized picture of
$t_{2g}$ electrons. 
\section*{Acknowledgements}
The authors gratefully thank Division of Instrumental Analysis, Okayama
University for the x-ray measurements, G. Radtke for the discussions
 about band calculations, and H. Nakao for useful discussions.
This work was partially supported by the Grant-in-Aid(No. 23110714).
One of authors(T. Igarashi) was supported by Program for Leading
 Graduate Schools" Integrative Graduate Education and Research in Green
 Natural Sciences", MEXT, Japan.

 \end{document}